# High Efficiency Carrier Multiplication in PbSe Nanocrystals: Implications for Solar Energy Conversion

R. D. Schaller and V. I. Klimov*

*Chemistry Division, C-PCS, Los Alamos National Laboratory, Los Alamos, NM 87545*

**Abstract**

We demonstrate for the first time that impact ionization (II) [the inverse of Auger recombination (AR)] occurs with very high efficiency in semiconductor nanocrystals (NCs). Interband optical excitation of PbSe NCs at low pump intensities, for which less than one exciton is initially generated per NC on average, results in the formation of two or more excitons (carrier multiplication) when pump photon energies are more than three times the NC band gap energy. Generation of multiexcitons from a single photon absorption event is observed to take place on an ultrafast (picosecond) timescale and occurs with up to 100% efficiency depending upon the excess energy of the absorbed photon. Efficient II in NCs can be used to considerably increase the power conversion efficiency of NC-based solar cells.

Solar power is an important source of clean, renewable energy [1, 2]. The maximum calculated thermodynamic conversion efficiency in solar cells is 43.9% under concentrated solar illumination. This calculation is based upon the assumption that absorption of an individual photon with energy above a semiconductor band gap ($E_g$) results in formation of a single exciton and that all photon energy in excess of $E_g$ is lost through electron-phonon interactions [3]. However, this apparent thermodynamic limit can be overcome once methods of harnessing the excess energy are developed.

Several methods have been offered to increase the power conversion efficiency of solar cells including the development of tandem cells, impurity band and intermediate band devices, hot electron extraction, and carrier multiplication [2, 4, 5]. Carrier multiplication, which was first observed in bulk semiconductors in the 1950s, would provide increased power conversion efficiency in the form of increased solar cell photocurrent. As was first proposed by Nozik, nanosize semiconductor crystals [nanocrystals (NCs)] might provide a regime where carrier multiplication could be greatly enhanced through impact ionization (II) [2, 6]. II is an Auger-type process whereby a high-energy exciton, created in a semiconductor by absorbing a photon of energy $\geq 2E_g$, relaxes to the band edge via energy transfer of at least $1E_g$ to a valence band electron, which is excited above the energy gap (Fig. 1(a)). The result of this energy transfer process is that *two excitons are formed for one absorbed photon*. Thus, this process converts more of the high photon energy portion of the solar spectrum into usable energy.

Here, for the first time, we demonstrate carrier multiplication via II in NCs. By directly monitoring exciton conversion to biexcitons in the time domain, we show that II in PbSe NCs is highly efficient, extremely fast, and occurs in a wavelength range that has potential to provide significantly increased solar cell power conversion efficiency.

Auger recombination (AR), the opposite of II, is a process in which an exciton recombines via energy transfer to an electron (or hole) that is excited to a higher energy state within or outside a NC (Fig. 1(b)). Because of restrictions imposed by energy and momentum conservation, AR is inefficient in bulk materials. However, AR becomes efficient in NCs due to enhanced Coulomb interactions and relaxation of momentum conservation [7-10]. Because of symmetry of the matrix elements that describe both II and AR, the former can also be very efficient in quantum-confined systems.

Here, we use transient absorption (TA) to monitor carrier population dynamics in high quality, oleic acid-passivated, colloidal PbSe NC samples [11] (size dispersity was ~5-10%, studied NC diameters were ~4 to 6 nm). Pump pulses (50 fs) from an amplified Ti-sapphire laser (pump photon energies, $\hbar\omega$ = 1.55 or 3.10 eV) or from a tunable optical parametric amplifier (OPA) excited NCs dissolved in hexane. The absorption change, $\Delta\alpha$, within the photo-excited spot is probed with 100 fs pulses that are tuned via another OPA to the band-edge ($A_1$) absorption maximum. As a measure of excitation density, we use an average number of photo-generated e-h pairs per NC, $N_{eh}$, produced by the pump pulse that we can accurately calculate and experimentally verify [9].

One important problem in experimental studies of II is to reliably detect this effect and quantify its efficiency. Prior to this work, II in solar cells had only been indirectly observed in conventional bulk silicon-based devices, however, with very low efficiencies at photon energies relevant to solar energy conversion [12]. Even with $E_g$ tuning via bulk Si-Ge alloys, the optimized material ($E_g$ = 0.83 eV) only resulted in a <1% improvement in solar power conversion efficiency [13]. Furthermore, studies of solar cell performance can only provide indirect information about II efficiency as other processes influence power conversion efficiency. In this report, we study II in NCs by directly monitoring the conversion of highly excited excitons into biexcitons. Our approach is to distinguish these two species via their relaxation dynamics. While excitons recombine slowly in PbSe NCs (with a sub-microsecond time constant [14]), biexcitons recombine very rapidly via AR on a picosecond timescale (see data below). Therefore, we are able to observe the generation of multiexcitons due to an easily discernable fast decay component in NC population dynamics. We are also able to accurately quantify the conversion efficiency of excitons to biexcitons ($\eta_{ii}$) by comparing the amplitude of the fast, multiexciton-related component with the height of the long-lived "excitonic" background as explained in Figs. 1(c-d).

To measure characteristic decay constants of biexcitons, we first study PbSe NC samples using 1.55 eV excitation for which II *cannot* occur ($\hbar\omega/E_g<2$ for each sample). As a measure of NC populations we use the normalized pump-induced bleaching of the lowest $A_1$ absorption maximum ($-\Delta\alpha/\alpha_0$; $\alpha_0$ is the absorption coefficient of the unexcited sample). As indicated in Fig. 2(a), absorption saturation measurements show that $\Delta\alpha$ is almost linear with pump intensity up to $N_{eh}\approx 3$, consistent with high 8-fold degeneracy of the lowest quantized states [15]. Therefore, $\Delta\alpha$ provides an accurate measure of $N_{eh}$ in both the single exciton and biexciton excitation regimes studied here. At low pump intensities, corresponding to the photo-generation of less than a single exciton per nanocrystal on average ($N_{eh}=0.6$) [9, 15, 16] (lower trace in Fig.2(b)), a bleach of the $A_1$ absorption maximum is observed that relaxes slowly (on a much longer timescale than our 2 ns delay stage can measure), consistent with slow, radiative recombination of single excitons [14]. As the pump intensity is increased to $N_{eh}=1.6$ (upper trace in Fig. 2(b)), the bleach magnitude becomes larger and a faster relaxation component (on the picosecond timescale) appears that can be attributed to AR of biexcitons. A small portion of this "biexcitonic" component is also discernible at $N_{eh}$=0.6 (lower trace in Fig. 2(b)), because, according to Poissonian statistics, there is also a finite probability to generate biexcitons, given by $P_2 = \frac{N_{eh}^2}{2}e^{-N_{eh}}$, even at $N_{eh}$<1. This fast component, measured using $N_{eh}$=1.6, can be isolated from the single exciton dynamics [9] and fits well to a single exponential decay (inset in Fig. 2(b)). The fast relaxation rates measured for NCs of different sizes are found to be inversely proportional to the NC volume (red squares in Fig. 2(c)), as was previously observed for AR in CdSe NCs [9]. This result confirms that the fast, pump intensity-dependent component originates from AR of biexcitons, which form as the result of absorption of two 1.55 eV photons by a NC.

Next, we performed TA studies of the same NC samples using 3.10 eV pump photons (Fig. 2(d)). For each of the studied NC samples, absorption of a 3.10 eV photon creates an exciton that has more than $3E_g$ (i.e. $\hbar\omega/E_g$>3) making II a possible relaxation route. Using low excitation intensities ($N_{eh}$=0.25) (determined taking into account the differences in absorption cross sections at the different photon excitation energies), we again measure $A_1$ band edge bleach dynamics. While at comparable excitation densities with 1.55 eV pump photon energy we primarily observed slow "excitonic" dynamics (upper trace Fig. 2(d)), we now clearly observe a large amplitude, fast component to the relaxation (lower trace in Fig. 2(d)) that is typical of biexciton dynamics. For all NC sizes studied with 3.10 eV excitation, this fast relaxation component decays with the same rate as that of the AR "biexcitonic" decay measured using the 1.55 eV pump photon energy (compare insets of Figs. 2(b) and 2(d)). Furthermore, the fast relaxation time constant (blue circles in Fig. 2(c)) follows the same dependence on NC volume as that observed using 1.55 eV pump photon energy (red squares in Fig. 2(c)), again indicative of AR. Further evidence that the detected biexcitons form due to II and not due to errors in pump intensity or absorption cross-section is apparent from the absolute magnitudes of changes in absorption (Fig. 2(d)).

We can calculate the efficiency of II in PbSe NCs using the procedure described in Fig. 1(d). As shown in Fig. 3(a-c), for a fixed pump photon energy (3.10 eV), the efficiency of II increases as $E_g$ decreases (Fig. 3(d)). As indicated in Fig. 3(d), the threshold for II observed in our experiments is close to $3E_g$. This observation can be explained by the "mirror" symmetry between the conduction and valence bands in PbSe [17], i.e., the excess energy of the absorbed pump photon is partitioned approximately equally between the electron and hole of a photo-generated exciton. Therefore, the energy conservation requirement for II can only be satisfied if either an electron or hole has an excess energy [approximately $(\hbar\omega-E_g)/2$] of at least $1E_g$. Studies performed using a pump pulse of tunable photon energy and a fixed NC $E_g$ (Fig. 3(e)) show a similar increase in II efficiency with excess energy as that observed for the case of a "variable" $E_g$ (Fig. 3(d)). Efficiencies as high as 118% are observed for photon energies of $3.8E_g$. Such a high efficiency indicates that some of the absorbed photons produce not just biexcitons but triexcitons, meaning that in some NCs *both* the electron and the hole undergo II.

Despite the tens to hundreds of picosecond lifetime of biexcitons in PbSe NCs, the II-generated excitons will be useful for solar power generation. It has previously been demonstrated for different solar cell systems that the charge transfer step can be very fast (~200 fs) and may occur with near unity efficiency [18, 19]. Therefore, AR should not compete with charge separation if the system is designed properly.

The theoretical treatment of Auger effects in NCs is extremely complex [7, 10], and therefore a theoretical picture of II in NCs has not been significantly developed. The conventional picture of II in bulk semiconductors is that exciton relaxation via II is competitive with intraband relaxation [2]. A quantum mechanical picture of resonance between degenerate wave functions, in which either 2 (a single exciton) or 4 carriers (a biexciton) exist, is also conceivable for NCs. Currently, we believe that the latter mechanism should produce a TA signal that is essentially instantaneous, while the former should result in a delayed signal. Inspection of the TA signal rise time (Fig. 3(b) inset) shows that a measurable build-up time (with a picosecond risetime) exists for TA signals that have a significant II component. Thus, II is very fast and competes with intraband relaxation in NCs, but does not appear to be an instantaneous process. Therefore, we believe that the process occurs via the more conventional picture of II, in which II competes with intraband relaxation. This conclusion is also consistent with our recent results on intraband relaxation, which show that intraband relaxation is slower than II, at least for the NC sizes studied here [20].

For a photovoltaic cell based on PbSe NCs of a single size, we can estimate the NC $E_g$ required to achieve maximum power conversion efficiency in the presence of II under concentrated solar illumination conditions, assuming the internal quantum efficiency of the device to be 1 (see Ref. 21). Shown in Fig. 4(a) are plots describing the effect of different II efficiencies for the case of a $3E_g$ onset of the process. The optimal $E_g$ shifts toward lower energy with increasing $\eta_{ii}$ and achieves a 10% increase in relative power conversion efficiency at $\eta_{ii}$=100% in comparison to the efficiency of a cell without II (48.3% vs. 43.9%) [3]. Further improvement in power conversion efficiency can be achieved by reducing the threshold for II. Shown in Fig. 4(b) are the power conversion efficiencies as a function of material $E_g$ for different onsets of II. A 37% increase in relative power conversion efficiency (to 60.3%) can be achieved via minimization of the II threshold to $2E_g$, which should be realizable in NCs of materials that have significantly different carrier effective masses.

Many properties of PbSe NCs, including efficient II, lend themselves well to high efficiency solar cells. PbSe NCs have a broadly size-tunable $E_g$ (~0.3 to 1.3 eV) that facilitates the construction of tandem cells, and they absorb strongly from the ultraviolet to the near-IR. Moreover, functional solar cells based upon semiconductor NCs have been demonstrated [22, 23]. Finally, II is also likely to provide significant benefits with regard to other desirable properties of NCs such as reduced pump thresholds in NC-based optical amplifiers [15], lasers [15, 24], and saturable absorbers as well as increased gain in avalanche photodiodes [25].

**Acknowledgements.** This work was supported by the Chemical Sciences, Biosciences, and Geosciences Division of the Office of Basic Energy Sciences, Office of Science, U.S. Department of Energy and Los Alamos LDRD funds.

**Figure Captions**

**Fig 1.** (a) II and (b) AR processes. Electrons (filled red circle), holes (empty red circle), conduction band (labeled **c**) and valence-band (labeled **v**). (c) Immediately following high photon-energy excitation ($\hbar\omega/E_g$>3 in our measurements), highly excited excitons form in some NCs. A fraction of these ($n_{xx}$) undergo II to create biexcitons, while others simply cool to the band edge, remaining as single excitons ($n_x$). Biexcitons that form via II next undergo AR to produce an exciton at long time. (d) Representative time-resolved data in which carrier populations are monitored in the pump intensity regime of $N_{eh}$<1 for low (red line) and high (blue line) pump photon energy (here below or above $3E_g$, respectively). The fast relaxation component in the blue trace is due to AR of biexcitons that have been generated via II. Its amplitude provides a direct measure of II efficiency: $\eta_{ii}=n_{xx}/(n_x+n_{xx})$=(A-B)/B.

**Fig 2.** (a) Photo-excitation at 1.55 eV produces a change of absorption ($\Delta\alpha$) at the $A_1$ absorption maximum (here 0.86 eV) that is linear up to $N_{eh}$≈3. (b) Carrier

relaxation dynamics monitored at the $A_1$ absorption feature (0.86 eV) using 1.55 eV photo-excitation, for which II *cannot occur*. At $N_{eh}$=0.6 (red line), primarily slow relaxation dynamics are observed. At $N_{eh}$=1.6 (black line), $\Delta\alpha$ becomes larger and a fast relaxation component becomes well pronounced; this component (shown in the inset after isolation from the slow component [9]) corresponds to rapid AR. (c) The lifetime of the fast relaxation component depends linearly upon the NC volume (red squares), which is indicative of AR of biexcitons. (d) A TA trace recorded with $N_{eh}$=0.25 using 3.10 eV pump photon energy (blue line), for which II is possible, shows a fast, "biexcitonic" relaxation component when monitoring the $A_1$ absorption feature. The extracted fast component (inset) is nearly identical to the inset shown in panel (b). For comparison, a trace of single exciton relaxation dynamics recorded using 1.55 eV photo-excitation with $N_{eh}$=0.6 has been replotted from panel (a) (red line). The NC-volume dependence of the relaxation constant of the fast component, observed in the II regime (blue circles in panel (c)) closely agrees with the biexciton lifetimes measured using 1.55 eV excitation (red circles in panel (c)).

**Fig 3.** (a-c) The $A_1$ feature relaxation dynamics, normalized at long time delay, measured for PbSe NCs having three different values of $E_g$ are shown using photo-excitation at 3.10 (blue) and 1.55 eV (red) with $N_{eh}$<0.5. These studies show that II efficiency is dependent upon the photo-generated exciton excess energy. The inset in panel (b) shows the measurable build-up of the TA signal for 3.10 eV excitation (blue line) and an autocorrelation of the pump pulse (black

line). (d) II efficiencies as a function of pump photon energy are compared for two cases: 1) the pump photon energy is fixed and $E_g$ of the NCs is changed (black squares), and 2) $E_g$ is fixed and the pump photon energy is changed (red circles). (e) For a fixed NC energy gap ($E_g$=0.94 eV), a tunable pump laser source was used to vary the pump photon energy and the efficiency of II was measured. All traces are normalized at long delay.

**Fig 4.** Calculated power conversion efficiencies of single stage solar cells as a function of material $E_g$ for different efficiencies and onsets of II. (a) For a 3$E_g$ II onset, a 10% increase (to 48.3%) in conversion efficiency is calculated for $\eta_{ii}$=100% (at $E_g$=0.8 eV) relative to the 43.9% efficiency that can be obtained without II ($\eta_{ii}$=0%). (b) The power conversion efficiency, shown for different onsets of II for $\eta_{ii}$=100%, can increase conversion efficiency by 37% (to 60.3%), if the onset is decreased to 2$E_g$.

*Schaller et al.*

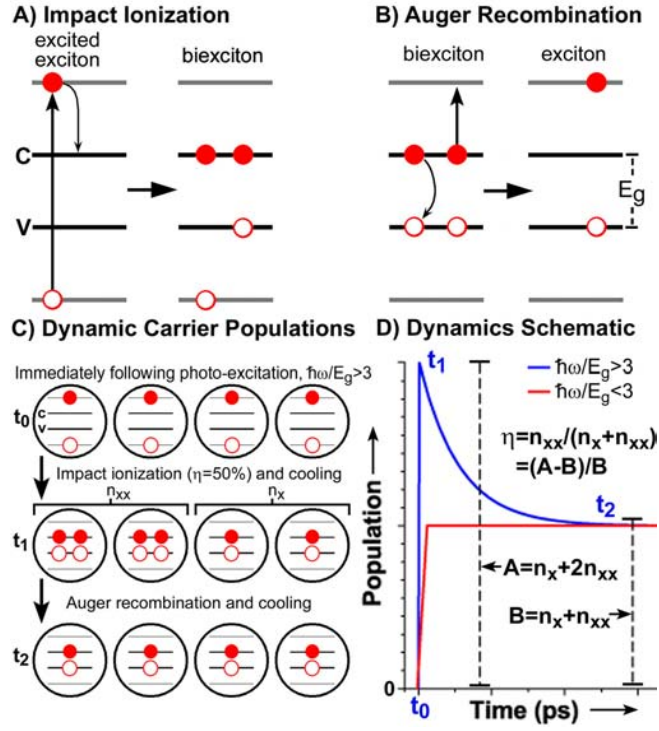

Figure 1

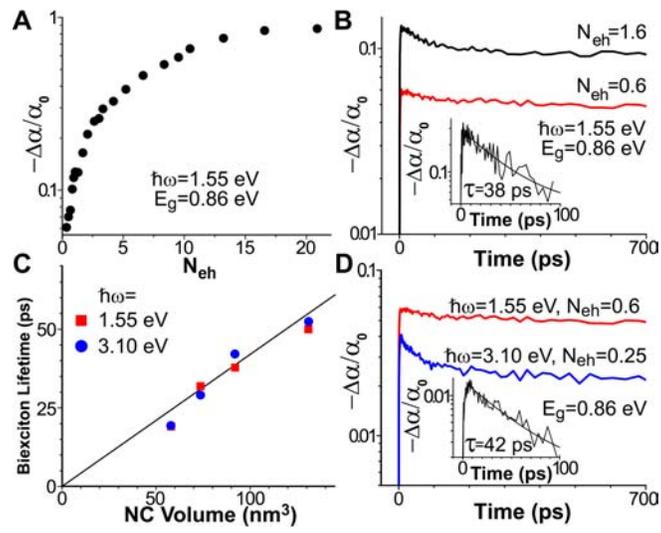

Figure 2

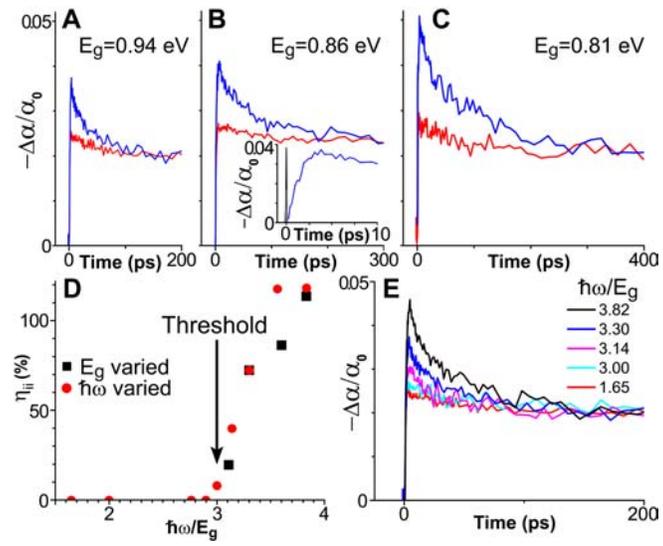

Figure 3

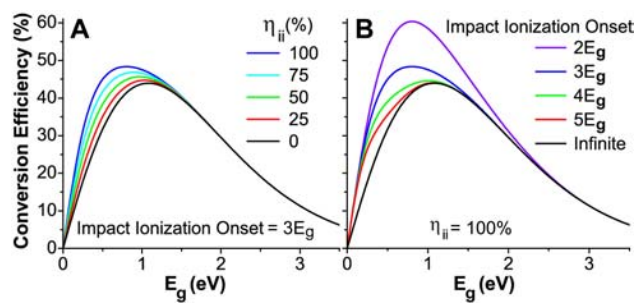

Figure 4